\documentclass[aps,prd,preprintnumbers,superscriptaddress,amsmath,amssymb,floatfix,longbibliography]{revtex4-2}
\pdfoutput=1
\usepackage[colorlinks,citecolor=blue,urlcolor=blue,linkcolor=red]{hyperref}
\usepackage{subfigure}
\usepackage{graphicx}
\usepackage{amsmath}
\usepackage{bm}

\begin{document}
	\title{Comment on ``Improved measurement of $\eta/\eta'$ mixing in $B_{(s)}^0\to J/\psi \eta^{(\prime)}$ decays"}
	
	%%%%%%%%%%%%%%%%%%%%%%%%%%%
	\author{Ying Chen}
	\email{cheny@ihep.ac.cn}
	\affiliation{Institute of High Energy Physics, Chinese Academy of Sciences, Beijing 100049, People's Republic of China}
	\affiliation{School of Physical Sciences, University of Chinese Academy of Sciences, Beijing 100049, People's Republic of China}
	\affiliation{Center for High Energy Physics, Henan Academy of Sciences, Zhengzhou 450046, People's Republic of China}
	
	\author{Geng Li}
	\email{ligeng@ihep.ac.cn}
	\affiliation{Institute of High Energy Physics, Chinese Academy of Sciences, Beijing 100049, People's Republic of China}
	\affiliation{Center for High Energy Physics, Henan Academy of Sciences, Zhengzhou 450046, People's Republic of China}
	%%%%%%%%%%%%%%%%%%%%%%%%%%%
	
	\begin{abstract}
		Instead of the glueball-$\eta'$ mixing ansatz, the latest measured ratios of the branching fractions of $B_{(s)}^0\to J/\psi \eta^{(\prime)}$ decays by LHCb can be understood by including the contribution from the light quark annihilation effect enhanced by the QCD $\mathrm{U}_A(1)$ anomaly for light pseudoscalar mesons. 
	\end{abstract}
	
	\maketitle
	
	%\section{Introduction}
	Very recently, the LHCb collaboration reported their improved measurement of the ratios of the branching fractions of the decay processes $B_{(s)}^0\to J/\psi \eta^{(\prime)}$~\cite{LHCb:2025sgp}. The precise results are 
	\begin{equation}\label{eq:data}
		\begin{aligned}
			&\frac{\mathcal{B}(B^0\to J/\psi \eta^\prime)}{\mathcal{B}(B^0\to J/\psi \eta)} = 0.48\pm 0.06,~~~
			\frac{\mathcal{B}(B^0\to J/\psi \eta)}{\mathcal{B}(B_s^0\to J/\psi \eta)} = (2.16\pm 0.18)\times 10^{-2},\\
			&\frac{\mathcal{B}(B_s^0\to J/\psi \eta^\prime)}{\mathcal{B}(B_s^0\to J/\psi \eta)} = 0.80\pm 0.03,~~~
			\frac{\mathcal{B}(B^0\to J/\psi \eta^\prime)}{\mathcal{B}(B_s^0\to J/\psi \eta^\prime)}=(1.33\pm 0.14)\times 10^{-2},  
		\end{aligned}
	\end{equation}
	where the errors are added in quadrature. By assuming the possible $\eta'$-glueball mixing, the authors of Ref.~\cite{LHCb:2025sgp} also use these results to contrain the $\eta/\eta'$ mixing angle $\phi_P$ and explore the glueball content ($|G\rangle$) of $\eta'$ through the flavor wave functions of $\eta$ and $\eta'$~\cite{DiDonato:2011kr,Fleischer:2011ib,Cheng:2008ss}, namely,
	\begin{equation}
		\begin{aligned}
			|\eta\rangle&= \cos\phi_P |\eta_l\rangle-\sin\phi_P |\eta_s\rangle,\\
			|\eta'\rangle&=\cos\phi_G\left(\sin\phi_P|\eta_l\rangle+\cos\phi_P|\eta_s\rangle\right)+\sin\phi_G |G\rangle,
		\end{aligned}
	\end{equation}
	where $|\eta_l\rangle$ and $|\eta_s\rangle$ have the quark configurations $(u\bar{u}+d\bar{d})/\sqrt{2}$ and $|s\bar{s}\rangle$ respectively, and $\sin\phi_G$ measures the glueball content of $\eta'$ ($\eta$ is mainly the flavor octet whose glueball content is temporarily neglected). In the limit of the flavor SU(3) syammetry, the following relations are obtained
	\begin{equation}\label{eq:ratio-ds}
		\begin{aligned}
			R_d&\equiv \frac{\mathcal{B}(B^0\to J/\psi \eta^\prime)}{\mathcal{B}(B^0\to J/\psi \eta)}\cdot \frac{\Phi^3(B^0\to J/\psi \eta)}{\Phi^3(B^0\to J/\psi \eta^\prime)}=\tan^2\phi_P \cos^2\phi_G,\\
			R_s&\equiv \frac{\mathcal{B}(B_s^0\to J/\psi \eta^\prime)}{\mathcal{B}(B_s^0\to J/\psi \eta)}\cdot \frac{\Phi^3(B_s^0\to J/\psi \eta)}{\Phi^3(B_s^0\to J/\psi \eta^\prime)}=\cot^2\phi_P \cos^2\phi_G,
		\end{aligned}
	\end{equation}
	and 
	\begin{equation}
		\begin{aligned}\label{eq:eta-etap}
			R_{\eta'}&\equiv \frac{\mathcal{B}(B^0\to J/\psi \eta')}{\mathcal{B}(B_s^0\to J/\psi \eta')}\cdot \frac{\Phi^3(B_s^0\to J/\psi \eta')}{\Phi^3(B^0\to J/\psi \eta')}=\frac{\tau_{B^0} m_{B^0}}{\tau_{B_s^0}m_{B_s^0}}\left|\frac{V_{cd}}{V_{cs}}\right|^2\frac{\tan^2\phi_P}{2},\\
			R_\eta&\equiv \frac{\mathcal{B}(B^0\to J/\psi \eta)}{\mathcal{B}(B_s^0\to J/\psi \eta)}\cdot \frac{\Phi^3(B_s^0\to J/\psi \eta)}{\Phi^3(B^0\to J/\psi \eta)}=\frac{\tau_{B^0} m_{B^0}}{\tau_{B_s^0}m_{B_s^0}}\left|\frac{V_{cd}}{V_{cs}}\right|^2\frac{\cot^2\phi_P}{2},
		\end{aligned}
	\end{equation}
	where $\tau$, $m$ and $V_{ij}$ are lifetime, mass and CKM matrix elements for specific states, respectively, and $\Phi(B^0_{(s)}\to J/\psi \eta^{(\prime)})$ is the two-body phase space factor
	\begin{equation}
		\Phi(B_{(s)}^0\to J/\psi \eta^{(\prime)})=\sqrt{\left[1-\left(\frac{m_{J/\psi}-m_{\eta(\prime)}}{m_{B_{(s)}^0}}\right)^2\right]\left[1-\left(\frac{m_{J/\psi}+m_{\eta^{(\prime)}}}{m_{B_{(s)}^0}}\right)^2\right]}.
	\end{equation}
	With the results in Eq.~(\ref{eq:data}) and PDG data~\cite{ParticleDataGroup:2024cfk}, the relations in Eq.~(\ref{eq:ratio-ds}) gives the very precise values of $\phi_P$ and $\phi_{G}$,
	\begin{equation}
		\phi_P=(41.6_{-1.2}^{+1.0})^\circ,\quad \phi_G=(28.1_{-4.0}^{+3.9})^\circ.
	\end{equation}
	The large $\eta'$-glueball mixing angle $\phi_G$ is surprising and may imply a light pseudoscalar glueball close to $\eta'$ in mass. However, the lattice QCD calculations in the quenched approximation predict the lowest pseudoscalar glueball has a mass of 2.4-2.6 GeV~\cite{Morningstar:1999rf,Chen:2005mg}, which is also supported by the lattice QCD studies with dynamical light quarks~\cite{Richards:2010ck,Gregory:2012hu,Sun:2017ipk}. There are also lattice QCD studies on the $\eta'$-glueball and $\eta_c$-glueball mixings which come out with very small mixing angles, namely, $|\theta_G|\approx 3.5(5)^\circ$ for the former~\cite{Jiang:2022ffl} and $|\theta_G|\approx 4.6(6)^\circ$ for the later (if $X(2370)$ is taken as a candidate for the pseudoscalar glueball)~\cite{Zhang:2021xvl}. The large difference between these values and $\phi_G\approx 28(4)^\circ$ cannot be reconciled easily. 
	\begin{figure}[t]
		\centering
		\subfigure[$CT$ diagram]{
			\includegraphics[width=0.4\linewidth]{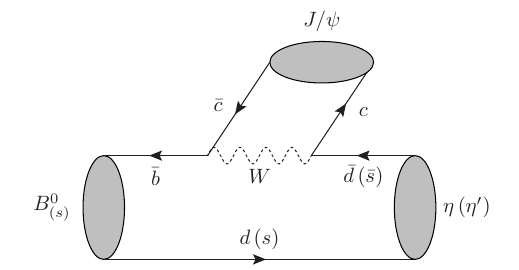}}
		\subfigure[$EA$ diagram]{
			\includegraphics[width=0.5\linewidth]{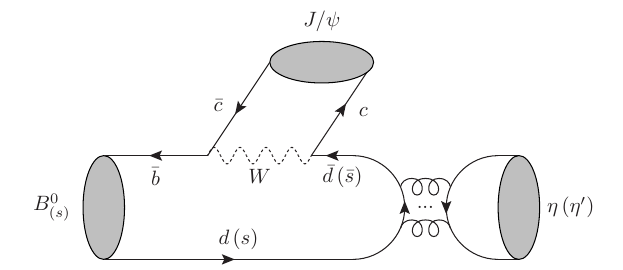}}
		\caption{The leading topologies contributing to the decays $B_{(s)}^0\to J/\psi \eta^{(\prime)}$. The left panel is the colour-suppressed tree diagram ($CT$) that dominates the decay rates, and the right panel is the $W$-exchange diagram ($EA$) where the coupling to $\eta^{(\prime)}$ is though gluonic interaction and can be enhanced by the QCD $\mathrm{U}_A(1)$ anomaly.}
		\label{fig:topology}
	\end{figure}
	
	Actually, the ratios in Eq.~(\ref{eq:data}) can be understood in an alternative way without $\eta'$-glueball mixing. The leading contributions to the decays $B_{(s)}^0\to J/\psi \eta^{(\prime)}$ are illustrated in Fig.~\ref{fig:topology} (There are also other contributions from penguin diagrams and penguin annihilation diagrams~\cite{Fleischer:2011au}, which are expected to be negligible). The left panel is the colour-suppressed tree diagram (CT) that is expected to dominate the decay rates. The right panel is the $W$-exchange diagram where the coupling to $\eta^{(\prime)}$ is though gluonic interaction and can be enhanced by the QCD $\mathrm{U}_A(1)$ anomaly. Let $A_{CT}$ be the contribution of $CT$ to the amplitude, and $A_{EA}$ the $EA$ contribution, then the decay amplitudes (without $\eta'$-glueball mixing) can be expressed as~\cite{Fleischer:2011ib} 
	\begin{equation}\label{eq:amp}
		\begin{aligned}
			\mathcal{A}(B_s^0\to J/\psi \eta)&\propto - F_{B_s} \sin\phi_P \left(A_{CT}-A_{EA}\right),~~~
			\mathcal{A}(B_s^0\to J/\psi \eta')\propto F_{B_s} \cos\phi_P \left(A_{CT}+2 A_{EA}\right),\\
			\mathcal{A}(B^0\to J/\psi \eta)&\propto F_{B} \cos\phi_P \left(A_{CT}+A_{EA}\right),~~~~~~
			\mathcal{A}(B^0\to J/\psi \eta')\propto F_{B} \sin\phi_P \left(A_{CT}+4 A_{EA}\right),
		\end{aligned}
	\end{equation}
	where $F_{B_{(s)}}$ is a factor irrelevant of the discussion in this comment. 
	
	Obviously, ignoring the EA contribution $A_{EA}$ and assuming the $\eta'$-glueball mixing one obtains the ratios in Eq.~(\ref{eq:ratio-ds}) and Eq.~(\ref{eq:eta-etap}). However, the assumption of $\eta'$-glueball mixing is not convincing, as addressed above. In constrast, the QCD $\mathrm{U}_A(1)$ anomaly may introduce important non-perturbative interaction between gluons and flavor singlet pseudoscalar mesons. First, the relatively heavy mass of $\eta'$ can be explained by this kind of interaction through the Witten-Veneziano machanism~\cite{Witten:1978bc,Witten:1979vv,Veneziano:1979ec}. Secondly, $\eta$ and $\eta'$ have large production 
	rates in the $J/\psi$ radiative decays where $\eta$ and $\eta'$ are generated purely through gluons. Recent lattice QCD studies also show that the production form factors of $\eta(\eta')$ in $J/\psi$ radiative decays are large~\cite{Jiang:2022gnd,Shi:2024fyv,Batelaan:2025vhx}. On the other hand, a pervious lattice QCD calculation come out with a production factor of pseudoscalar glueball in the $J/\psi$ radiative decay that is comparable with those of light pesudoscalar mesons~\cite{Gui:2019dtm}. Furthermore, the scalar form factor $f_0$ for the semileptionic decays $D_s\to \eta^{(\prime)}\bar{l}\nu_l$ has been calculated in lattice QCD~\cite{Bali:2014pva}, where a large contribution from the disconnected diagram (similar to $EA$ and corresponding to the gluonic production of $\eta'$) 
	is observed to be important for $D_s\to \eta'$. 
	These facts manifest that the usual OZI rule for the production of $\eta$ states is violated due to the $\mathrm{U}_A(1)$ anomaly. Therefore, if we discard the $\eta'$-glueball mixing ansatz but take the $EA$ contribution $A_{EA}$ into account, then the ratios in Eq.~(\ref{eq:ratio-ds}) and Eq.~(\ref{eq:eta-etap}) are modified as 
	\begin{equation}\label{eq:modified}
		\begin{aligned}
			R_d&= \tan^2\phi_P\frac{(1+4\alpha)^2}{(1+\alpha)^2},~~~
			R_\eta=\frac{\tau_{B^0} m_{B^0}}{\tau_{B_s^0}m_{B_s^0}}\left|\frac{V_{cd}}{V_{cs}}\right|^2\frac{\cot^2\phi_P}{2} \frac{(1+\alpha)^2}{(1-\alpha)^2},\\
			R_s&=\cot^2\phi_P \frac{(1+2\alpha)^2}{(1-\alpha)^2},~~~
			R_{\eta^\prime}=\frac{\tau_{B^0} m_{B^0}}{\tau_{B_s^0}m_{B_s^0}}\left|\frac{V_{cd}}{V_{cs}}\right|^2\frac{\tan^2\phi_P}{2} \frac{(1+4\alpha)^2}{(1+2\alpha)^2},
		\end{aligned}
	\end{equation}
	where $\alpha=A_{EA}/A_{CT}$ is assumed to be real. Then using the LHCb values of $R_{d,s}$ we obtain $\alpha=-0.039(10)$ from $R_d\cdot R_s$. This negative value 
	signals opposite signs of $A_{CT}$ and $A_{EA}$, which can be understood in two-folds: First, the $CT$ diagram in Fig.~\ref{fig:topology} involves a single quark loop, while the $EA$ diagram has two quark loops, such that an addition quark loop contributes an additional minus sign. Secondly, the lattice QCD study in Ref.~\cite{Bali:2014pva} also observes this phenomnon in the matrix elements for the semileptonic $D_s\to\eta^\prime$ decay. With this $\alpha$, the amplitudes in Eq.~(\ref{eq:amp}) imply that, the $EA$ contribution increases(decreases) the decay rate of $B^0_s\to J/\psi\eta$ ($B^0\to J/\psi \eta$) by roughly 8\%, and decreases 
	the decay rates of $B^0_s\to J/\psi\eta'$ and $B^0\to J/\psi \eta'$ by roughly 16\% and 32\%, respectively. In other words, the $EA$ contribution is very important for $B_{(s)}^0\to J/\psi \eta'$ decays. In the meantime, one can extract the values of the $\eta-\eta'$ mixing angle $\phi_P$ from each ratio of $R_{d,s,\eta,\eta^\prime}$ as following
	\begin{equation}
		\phi_P(R_d/R_s)=41.6(1.2)^\circ,\quad \phi_P(R_\eta)=44.2(1.4)^\circ, \quad \phi_P(R_{\eta^\prime})=39.5(2.2)^\circ.
	\end{equation}
	The value of $\phi_P(R_d/R_s)$ derived by our model is the same as that from Eq.~(\ref{eq:ratio-ds}) since $\phi_G$ terms are canceled here. The values of $\phi_P$ from $R_{\eta,\eta'}$ are a little different but consistent with $\phi_P(R_{d,s})$ within one-sigma. It should be noted that these values converge more than those from Eq.~(\ref{eq:eta-etap}) which gives $\phi_P(R_\eta)=46.9(6)^\circ$ and $\phi_P(R_{\eta^\prime})=36.6(7)^\circ$ in Ref.~\cite{LHCb:2025sgp}. This manifests that the inclusion of the $EA$ contribution is in the right trend. 
	
	To summarize, instead of the $\eta'$-glueball mixing ansatz adopted by LHCb, we interpret the ratios of the branching fractions of $B_{(s)}^0\to J/\psi \eta^{(\prime)}$ decays by including the light quark annihilation effect which can be important for the strong interaction production of light pseudoscalar mesons ($\eta$ and $\eta'$ in this comment) because of enhancement from the QCD $\mathrm{U}_A(1)$ anomaly. The amplitude ratio of this annihilation effect to the dominant mechanism (tree-level color suppression process) is determined to be $\alpha=-0.039(10)$, which implies that the annihilation effect decreases the decay rates of $B_{(s)}^0\to J/\psi \eta'$ by roughly 32\% (16\%). Using this $\alpha$ value, the values of the $\eta-\eta'$ mixing angle $\phi_P$ determined from different ratios $R_{d,s,\eta,\eta'}$ agree with each other better than without considering the annihilation effect. 
	
	\section*{Acknowledgements} 
	This work is supported by the National Natural Science Foundation of China (NNSFC) under Grants No.~12293060, No.~12293065 and No.~11935017. GL is also supported by China Postdoctoral Science Foundation under Grant No.~2025M773362.
	
	\bibliography{reference}

\end{document}